# Tunable Fe-vacancy disorder-order transition in FeSe thin films

Y. Fang[1], D. H. Xie[1], W. Zhang[1], F. Chen[2], W. Feng[1], B. P. Xie[2], D. L. Feng[2], X. C. Lai[1*] and S. Y. Tan[1*]

[1]Science and Technology on Surface Physics and Chemistry Laboratory, Mianyang 621908, China

[2]Physics Department, Applied Surface Physics State Key Laboratory, and Advanced Materials Laboratory, Fudan University, Shanghai 200433, China

Various Fe-vacancy orders have been reported in tetragonal $\beta$-$Fe_{1-x}Se$ single crystals and nanowires/nanosheets, which are similar to those found in alkali metal intercalated $A_{1-x}Fe_{2-y}Se_2$ superconductors. Here we report the *in-situ* angle-resolved photoemission spectroscopy study of Fe-vacancy disordered and ordered phases in FeSe multi-layer thin films grown by molecular beam epitaxy. Low temperature annealed FeSe films are identified to be Fe-vacancy disordered phase and electron doped. Further long-time low temperature anneal can change the Fe-vacancy disordered phase to ordered phase, which is found to be semiconductor/insulator with $\sqrt{5}\times\sqrt{5}$ superstructure and can be reversely changed to disordered phase with high temperature anneal. Our results reveal that the disorder-order transition in FeSe thin films can be simply tuned by vacuum anneal and the $\sqrt{5}\times\sqrt{5}$ Fe-vacancy ordered phase is more likely the parent phase of FeSe.

## I. INTRODUCTION

High superconducting transition temperature($T_c$) in FeSe and its intercalated compounds have attracted a great deal of attentions in the area of condensed matter physics. $\beta$-$Fe_{1+\delta}Se$ has the simplest chemical and crystal structure with only superconducting FeSe layers in iron-based superconductors, which undergoes tetragonal to orthorhombic structural transition at $T_s$~90 K and remarkable enhancement of $T_c$ from ~8 to 36.7 K under pressure at 8.9 GPa[1-3]. Intercalating alkali metal ions and small molecules in FeSe can also enhance Tc higher than 40 K[4-7]. Moreover, superconductivity in single-layer FeSe (SL-FeSe) thin films grown on $SrTiO_3$(STO) substrate is particularly fascinating. A superconducting gap as large as 20 meV was first discovered by scanning tunneling spectroscopy (STS), which was later confirmed by angle-resolved photoemission spectroscopy (ARPES) measurements[8-11]. Then, the $T_C$ above 40 K and 100 K in SL-FeSe films has been demonstrated by direct transport measurements[12] and *in-situ* electrical transport measurements[13], respectively. The FeSe and related compounds provide an ideal platform to study the mechanism of superconductivity in the iron-based superconductors.

In general, the parent compound which can be tuned to the superconductor by chemical doping or external pressure is of great importance to investigate the superconducting mechanism. The parent phase of the iron-pnictide superconductor is suggested to be bad metal with spin-density-wave(SDW) order. In $A_{1-x}Fe_{2-y}Se_2$ superconductor, there exists an intrinsic mesoscopic phase separation of a superconducting phase and an Fe-vacancy ordered antiferromagnetic insulating phase[14-16]. Antiferromagnetic insulating phase $K_2Fe_4Se_5$ with $\sqrt{5}\times\sqrt{5}$ Fe-vacancy order is most frequently observed Fe-vacancy order，and is suggested to be the parent compound owing to the close connection with superconducting phase[17,18]. However, the phase separation makes it intricate to experimentally determine the intrinsic nature of the superconducting phase. A recent study on the non-superconducting $Fe_4Se_5$[19], which exhibits the Fe-vacancy order and is magnetic, shows that it becomes superconducting after high temperature anneal, meanwhile the Fe-vacancy order disappears. This discovery suggests that the rich-phases found in

$A_{1-x}Fe_{2-y}Se_2$ are not exclusive in FeSe related superconductors, and the Fe-vacancy disorder and order have intimate relationship with superconductivity. To better understand the superconducting mechanism in FeSe, the experimental study on the electronic structure of Fe-vacancy disorder and order phases becomes essential.

In this paper, we report the *in-situ* angle-resolved photoemission spectroscopy study of Fe-vacancy disorder-order transition in FeSe multi-layer thin films grown by molecular beam epitaxy. High temperature(550℃ for 4 h) annealed multi-layer FeSe thin films are found to be metallic with nematic order. Low temperature(550℃ for 1 h and 250℃ for 3 h) annealed FeSe films are identified to be Fe-vacancy disordered phase with electron doping. Further long-time low temperature(250℃ for 16 h) anneal can change the Fe-vacancy disordered phase to ordered phase, which is found to be semiconductor /insulator with $\sqrt{5} \times \sqrt{5}$ superstructure and can be reversely changed to disordered phase with high temperature annealing. Our results reveal that the disorder-order transition in FeSe thin films can be simply tuned by vacuum anneal and the $\sqrt{5} \times \sqrt{5}$ Fe-vacancy ordered phase is more likely the parent phase of FeSe.

## II. EXPERIMENT

High-quality FeSe single crystalline thin films were grown on the $TiO_2$ terminated and Nb-doped $SrTiO_3$ (0.5%wt) substrate with the molecular beam epitaxy (MBE) method following the previous reports[8,11]. The STO substrate was degassed at 550℃ for 3 h in vacuum, and then heated to 950℃ under the Se flux for 30 min. During growth, the substrate was kept at 450℃ with the Se flux twenty times greater than Fe by co-deposition. The properties of FeSe thin films can be tuned by varying the anneal temperature and length of the anneal time. The film annealed at 550℃ after growth in vacuum for 4 h is referred as nematic multi-layer FeSe (NM-FeSe) hereafter. When the as grown film was annealed at 550℃ for 1 h and 250℃ for 3 h, we got the Fe-vacancy disordered FeSe ( referred as VD-FeSe hereafter). If we further anneal the VD-FeSe at 250℃ for 16 hours, the VD-FeSe changes to Fe-vacancy ordered FeSe( referred as VO-FeSe hereafter). The cobalt-doped FeSe(CD-FeSe) is induced by depositing cobalt atoms on the surface of as grown FeSe thin film and a subsequent 4 h of 550℃ annealing process[20].

After growth and anneal, the film was directly transferred from the MBE chamber into the ARPES chamber with typical vacuum of $5 \times 10^{-11}$ mbar. ARPES was conducted with 21.2 eV photons from a helium discharge lamp. A SCIENTA R4000 analyzer was used to record ARPES spectra with typical energy and angular resolutions of 10 meV and 0.2°, respectively. A freshly evaporated gold sample in electrical contact with the FeSe sample served to calibrate $E_F$.

## III. RESULTS AND DISCUSSIONS

The multi-layer FeSe thin films annealed at 550℃ for 4 hours have been demonstrated to have nematic order (NM-FeSe) in our previous studies[11,20]. The Fermi surface topologies and band structures of FeSe films thicker than 2 unit-cell(UC) are very much alike. Taking 50 UC FeSe film as an example, the observed Fermi surface of NM-FeSe consists of cross-like electron pockets centered at M and oval-shaped hole pockets centered at Γ, as shown in Figs.1(a1)and 1(b1). When the as grown multi-layer FeSe is annealed at 550℃ for 1 h and 250℃ for 3 h, the electronic structure changes dramatically(VD-FeSe). The Fermi surface of 40 UC VD-FeSe consists of oval-shaped electron pockets centered around M and some spectral weights at Γ, which is obviously electron doped with 0.06 $e^-$/Fe based on the calculated Luttinger volume[Figs.1(a4) and 1(b4)]. The Fermi surface of VD-FeSe is very similar to single-layer FeSe (SL-FeSe) and cobalt doped multi-layer FeSe(CD-FeSe), which are both heavily electron doped. The SL-FeSe have only circular electron pockets around M with typical 0.12 $e^-$/Fe [Figs.1(a2) and 1(b2)], which is thought to be doped by the electrons from STO substrate. The cobalt doped CD-FeSe also have circular electron pockets at M and some spectral weights at Γ with 0.08 e⁻/Fe [Figs.1(a3) and 1(b3)].

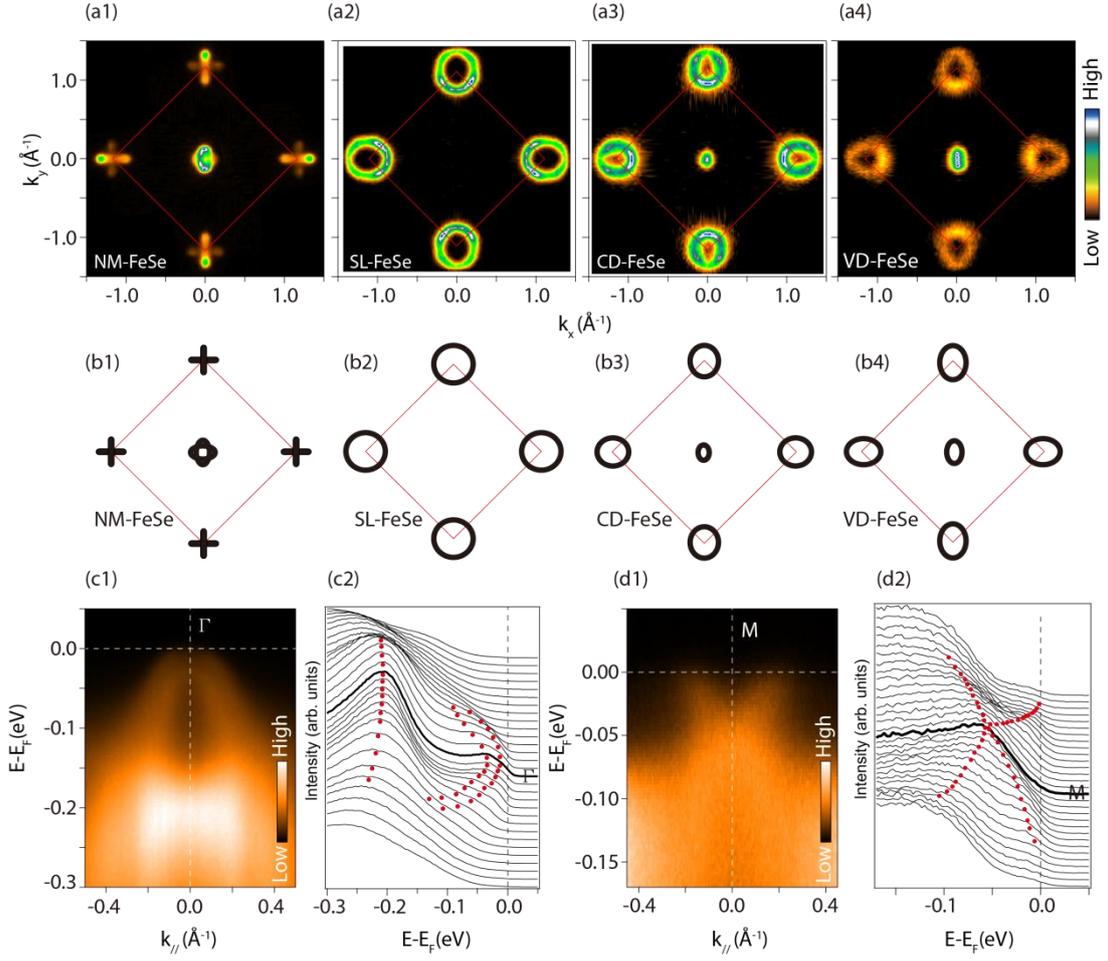

FIG.1. (Color online) The electronic structure of VD-FeSe. (a1-a4) Fermi surface topologies of various phases of FeSe thin films measured at 30 K integrated over a [$E_F$-10 meV, $E_F$+10 meV] window. The NM-FeSe, SL-FeSe, CD-FeSe and VD-FeSe stand for nematic multi-layer FeSe, single-layer FeSe, cobalt doped multi-layer FeSe and Fe-vacancy disordered multi-layer FeSe, respectively. (b1-b4) Schematic drawing of the Fermi surface topologies for various phases of FeSe thin films, the sizes of the Fermi pockets are extracted from (a) for clarity. (c1-c2) Photoemission intensity around Γ and the corresponding energy distribution curves(EDCs) of VD-FeSe. (d1-d2) Photoemission intensity around M and the corresponding energy distribution curves(EDCs) of VD-FeSe.

According to the *in-situ* STM measurements[21], the as-grown FeSe thin films have extra Se atoms on the surface due to the Se-rich growth condition. Annealing the as grown films at high temperature(typically 550℃ in our study) will remove the extra Se to form stoichiometric FeSe. The stoichiometric multi-layer FeSe have been demonstrated to be charge neutral and have nematic order in the previous studies[11]. The VD-FeSe films were annealed at a much lower temperature(250℃) compared with the NM-FeSe, which can't remove the extra Se completely and leave a Se rich(or Fe deficiency) surface. The Fe deficiency FeSe film is referred as Fe-vacancy disordered phase(VD-FeSe) in our paper. The band structure for VD-FeSe around Γ and M are shown in Figs.1 (c) and 1(d). Two hole-like bands can be clearly resolved at Γ, whose band tops barely touch $E_F$ and contribute to the spectral weights at the Brillouin zone center. At M, one electron-like band which forms the oval-shaped electron pocket and a hole-like band located beneath the electron band can be resolved.

It is a bit confusing that the Fe-vacancy disorder can cause electron doping to FeSe. A theoretical study[18] on $K_{1-x}Fe_{2-y}Se_2$ found that the Fe-vacancy disorder raises the chemical potential significantly, eliminating the hole pockets around Γ and enlarging the electron pockets at M. Berlijn.et al[22]studied the doping effects of Se vacancies in monolayer FeSe and found that in terms of the Fe-3d bands, Se vacancies (excess Fe) behave like hole dopants rather than electron dopants. However, a recent first-principles calculations[23] revealed that the excess Fe contributes electrons to the monolayer Fe-d states that may suppress the hole pockets observed around Γ, while excess Se hybridizes with the monolayer Fe-d states and partially opens a gap just above the Fermi energy. Our ARPES results support the former conclusions[18,22] that the Fe vacancies behave like electron dopants to FeSe, leading to the Fe-vacancy disordered FeSe with electron doping .

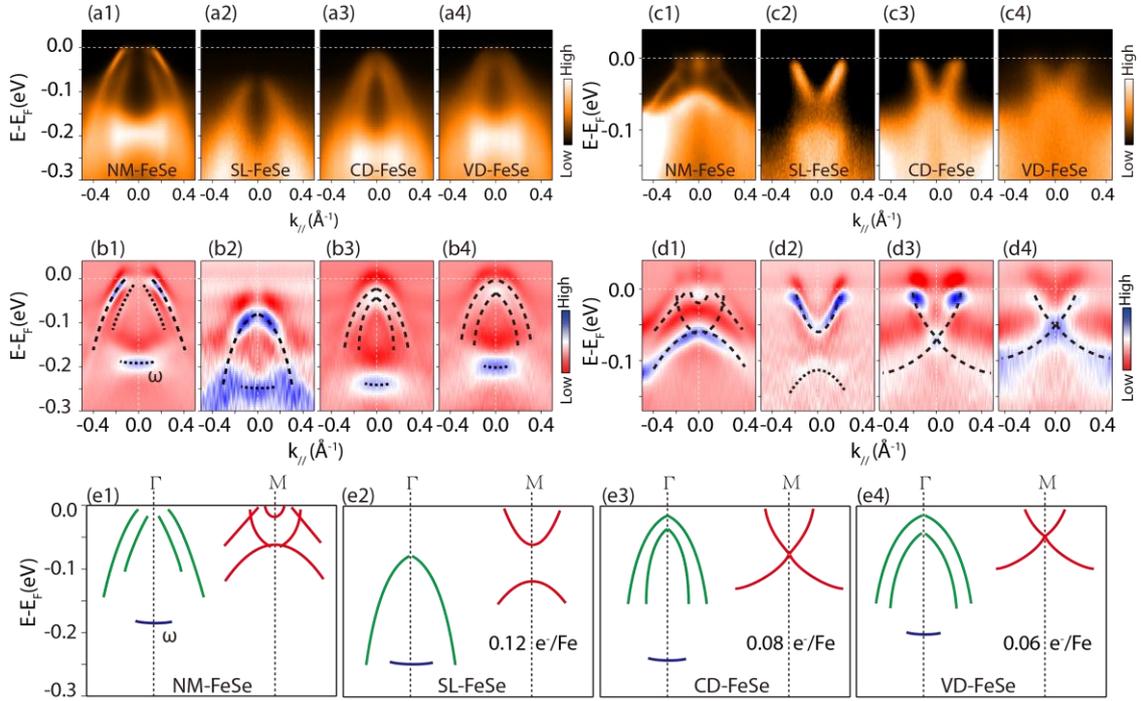

FIG.2. (Color online) Band structures of various phases of FeSe thin films. (a, b) The photoemission intensity (a1-a4) and its corresponding second derivative of the intensity plot(b1-b4) around Γ for NM-FeSe, SL-FeSe, CD-FeSe and VD-FeSe, respectively. (c, d) The photoemission intensity (c1-c4) and its corresponding second derivative of the intensity plot(d1-d4) around M for NM-FeSe, SL-FeSe, CD-FeSe and VD-FeSe, respectively. The dashed lines in (b) and (d) are guides to the eye. （e1-e4） Schematic band structures of NM-FeSe, SL-FeSe, CD-FeSe and VD-FeSe, respectively. The doping levels are calculated based on the Luttinger volume.

To better understand the electronic structure of the electron doped VD-FeSe, we present the low-energy band structures of NM-FeSe, SL-FeSe, CD-FeSe and VD-FeSe around Γ and M in Fig.2 for comparison. The NM-FeSe represent the pristine stage of undoped FeSe and all the other three are electron doped with different doping levels. From Fig.2(b) we can see that, the two hole-like bands which cross $E_F$ and a relatively flat band ω located at about -190 meV at Γ in NM-FeSe both move downwards to high binding energy upon electron doping. The larger the doping level is, the higher binding energy the bands move to. While there is only subtle difference between cobalt doped CD-FeSe and Fe-vacancy doped VD-FeSe at Γ, the two hole-like bands get totally reconstructed to one in SL-FeSe due to the interface effect. The obvious difference between the two electron doped FeSe without interface effect is that the effective mass of the hole bands are much smaller in CD-FeSe than in VD-FeSe, which indicates weaker correlation strength in CD-FeSe due to cobalt dosing.

The complicated band structure at M in NM-FeSe changes significantly with electron doping, as shown in Figs.2(c) and 2(d). Two electron-like bands and two hole-like bands exist at M in NM-FeSe, the electron doping suppress the nematic order and change the band structure topology. The electronic structure of CD-FeSe and VD-FeSe seems much alike, mainly consists of an electron band which crossing $E_F$ and a hole band located beneath. The bottom of the electron band located at -70 meV and -50 meV in CD-FeSe and VD-FeSe, respectively, and again the effective mass of the electron band in CD-FeSe is much smaller due to cobalt dosing. A gap of 60 meV opens between the intersect electron and hole bands in SL-FeSe due to interface effect, as shown in Fig.2(d2). The electronic structure of SL-FeSe is much different from the other two electron doped ones[Fig.2(e)], which indicates that the unique electronic structure of SL-FeSe is not originated from simply doping the multi-layer FeSe films with electron carriers, the FeSe/STO interface plays an important role for determination of the unique electronic structure.

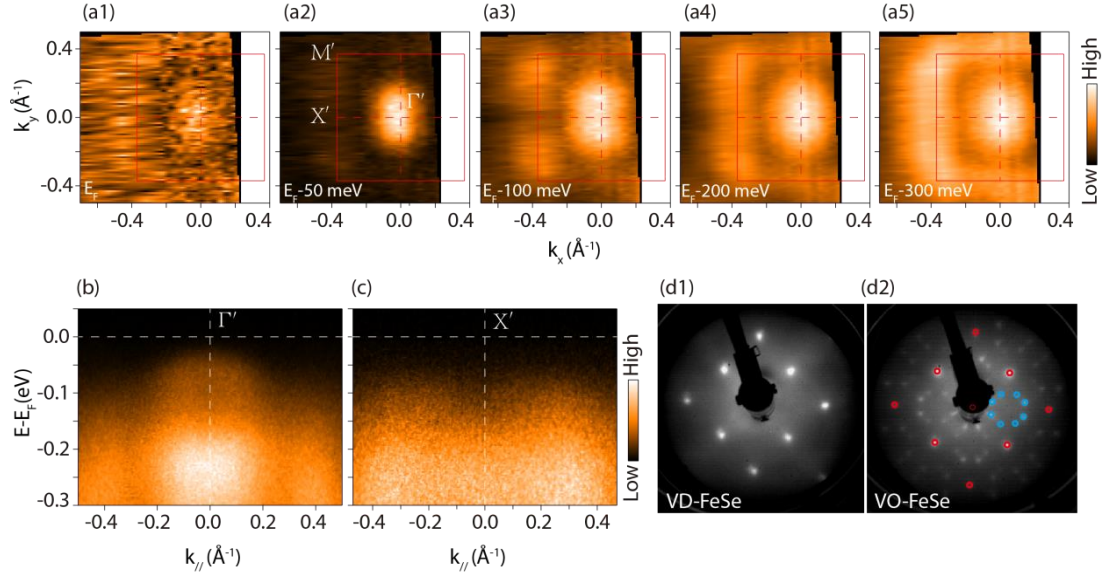

Fig.3. (Color online) The electronic structure of Fe-vacancy ordered FeSe.（a1-a5）Constant-energy maps at different binding energies from $E_F$ to $E_F$-300 meV for Fe-vacancy ordered FeSe.（b, c）The photoemission intensity of Fe-vacancy ordered FeSe around Γ and M, respectively.（d）The LEED patterns of VD-FeSe (d1) and VO-FeSe(d2).

When the Fe-vacancy disordered FeSe(VD-FeSe) was further annealed at 250℃ for 16 h, we got the Fe-vacancy ordered FeSe(VO-FeSe), which turns out be a semiconductor/insulator. The photoemission intensity maps of VO-FeSe are shown in Fig.3(a), which are integrated over a [$E_F$-20 meV, $E_F$+20 meV] window. As shown in Fig.3(a1), there is no obvious spectral weight at $E_F$. With decreasing binding energy, the spectral weight gets stronger and clear constant energy dispersion contour emerges at $E_F$-200 meV [Fig.3(a4)]. Large spectral weight and a rectangular shaped pocket appear around the Brillouin zone center, as shown in Figs.3(a4)and 3(a5). The Brillouin zone size of VO-FeSe determined from the high-symmetry points of the photoemission map is much smaller than VD-FeSe. The value of the lattice constant derived from the reciprocal of the Brillouin zone size is about a=b=8.5Å in VO-FeSe, which is $\sqrt{5}$ times of that in

VD-FeSe(a=b=3.8 Å).

The low-energy electron diffraction(LEED) patterns of VD-FeSe and VO-FeSe are shown in Fig.3(d). The LEED pattern of VD-FeSe shows typical tetragonal symmetry without any sign of superstructure [Fig.3(d1)]. Compared with the LEED pattern of VD-FeSe, we can see clear $\sqrt{5} \times \sqrt{5}$ superstructure in Fig.3(d2). The eight blue dots form a circular ring in between each four red dots, which is a clear evidence that $\sqrt{5} \times \sqrt{5}$ superstructures emerge on the surface. The band structure at the reconstructed high symmetry points Γ' and M' of VO-FeSe are shown in Figs.3(b) and 3(c). The density of states near $E_F$ are significantly suppressed at Γ' and M' in VO-FeSe, which shows typical semiconductor/insulator behavior. Judging from the ARPES results and the LEED pattern, we can conclude that the VO-FeSe is a Fe-vacancy $\sqrt{5} \times \sqrt{5}$ ordered semiconductor/insulator.

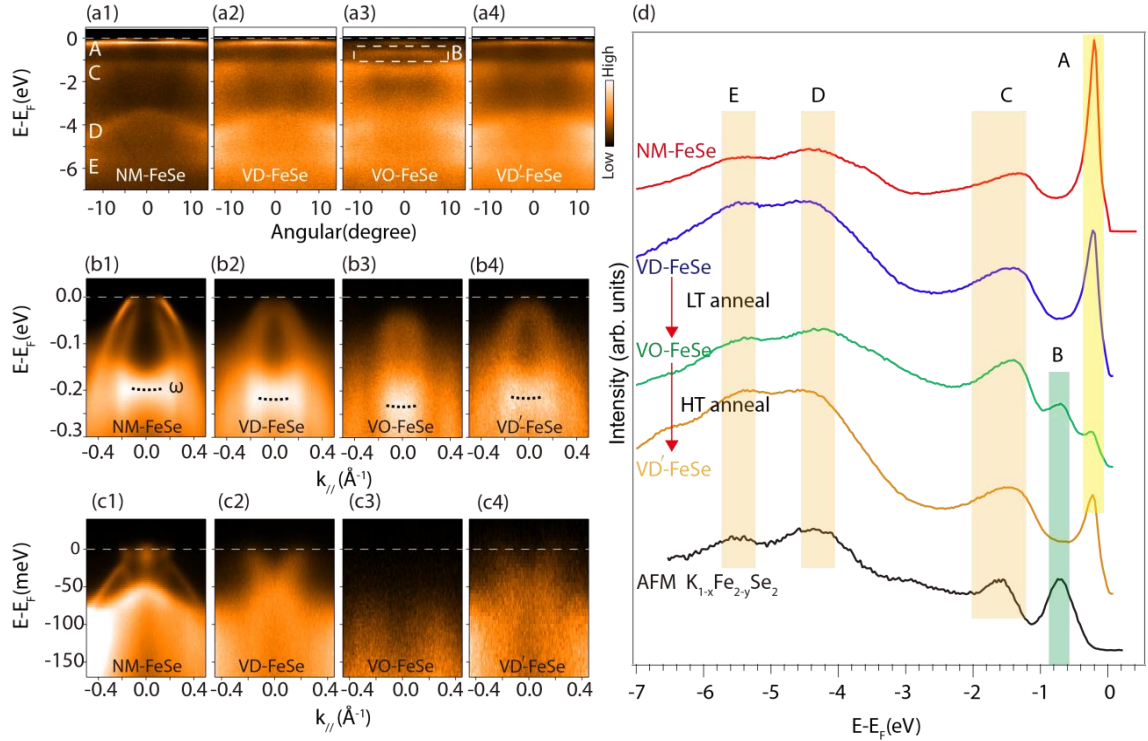

Fig.4. (Color online) The Fe-vacancy disorder-order transition in FeSe thin films. (a1-a4) The valence band structure a Γ for NM-FeSe, VD-FeSe, VO-FeSe and VD'-FeSe, respectively. (b1-b4) The low-energy band structure at Γ for NM-FeSe, VD-FeSe, VO-FeSe and VD'-FeSe, respectively. (c1-c4) The low-energy band structure at M for NM-FeSe, VD-FeSe, VO-FeSe and VD'-FeSe, respectively. (d) The angle integrated spectrum at Γ for NM-FeSe, VD-FeSe, VO-FeSe, VD'-FeSe and AFM phase of $K_{1-x}Fe_{2-y}Se_2$, respectively.

Interestingly, the Fe-vacancy ordered FeSe (VO-FeSe) can be reversely tuned to Fe-vacancy disordered FeSe (VD-FeSe) through 550℃ high temperature anneal. As mentioned above, the VD-FeSe changes to Fe-vacancy $\sqrt{5} \times \sqrt{5}$ ordered phase (VO-FeSe) under 250℃ anneal for 16 h. If we anneal the VO-FeSe at 550℃ for 2 h, the Fe-vacancy order will be destroyed and the film changes to Fe-vacancy disordered phase again(referred as VD'-FeSe). The valence band and low-energy band structure for NM-FeSe, VD-FeSe, VO-FeSe and VD'-FeSe are present in Fig.4. If we ignore the influence of sample quality on the band structure, the VD-FeSe and VD'-FeSe have almost the same characteristic band structures, as shown in Figs. 4(a2) and 4(a4), 4(b2) and 4(b4), 4(c2) and 4(c4). While the low-energy band structures have much difference between NM-FeSe, VD-FeSe and VO-FeSe[Fig4.(b),(c) and (d)], their valence bands are relatively the same[Fig.4(a)]. There are four main features in the valence bands for various phases of FeSe, which locate at about -0.2, -1.3, -4.4 and -5.5eV and are named A, C, D, E, respectively. A special band(band B) appears in the VO-FeSe, which lies at about -700 meV below $E_F$[Fig.4(a3)]. Band B can only be seen in VO-FeSe, which is the characteristic feature of the Fe-vacancy $\sqrt{5} \times \sqrt{5}$ ordered phase and has been found in antiferromagnetic(AFM) phase of $K_{1-x}Fe_{2-y}Se_2$ before[15].

We plot the angle integrated spectra at Γ for NM-FeSe, VD-FeSe, VO-FeSe, VD'-FeSe and AFM phase of $K_{1-x}Fe_{2-y}Se_2$ together in Fig.4(d). The C, D and E peaks can be seen in all the FeSe thin films and $K_{1-x}Fe_{2-y}Se_2$ single crystals, which may correspond to the characteristic features of FeSe related compounds. Peak B can only be seen in VO-FeSe(-700 meV) and AFM phase of $K_{1-x}Fe_{2-y}Se_2$(-710 meV), which corresponds to the Fe-vacancy $\sqrt{5} \times \sqrt{5}$ ordered semiconductor/ insulator phase. Peak A corresponds to the ω band in Fig.4(b), which appears in all the FeSe thin films. We speculate that the Fe-vacancy $\sqrt{5} \times \sqrt{5}$ ordered FeSe is a semiconductor/insulator similar to the insulator phase of $K_{1-x}Fe_{2-y}Se_2$, and there are two explanation that why we can see peak A in VO-FeSe. Firstly, only the upper most layer of VO-FeSe is Fe-vacancy $\sqrt{5} \times \sqrt{5}$ ordered and insulating. Because the probing depth of our ARPES measurement is about one to two atomic layers, we get the contribution of peak A from the interfacial layer. Secondly, there is mesoscopic phase separation in the VO-FeSe thin films which is similar to $K_{1-x}Fe_{2-y}Se_2$. Our ARPES results reveal that there is Fe-vacancy $\sqrt{5} \times \sqrt{5}$ ordered insulating/semiconducting phase in FeSe thin films and the disorder-order transition can be simply tuned by vacuum anneal, which is similar to $K_{1-x}Fe_{2-y}Se_2$.

## IV. DISCUSSION AND CONCLUSION

Identifying the parent compound and understanding its doping evolution is a crucial step to elucidate the superconducting pairing mechanism. For the potassium-intercalated FeSe compounds, there has been significant debate regarding what the parent compound is[24-27]. A recent study[28] shows that the Fe-vacancy ordered $K_2Fe_4Se_5$ is the magnetic, Mott insulating parent compound of the superconducting state. Insulating $K_2Fe_4Se_5$ becomes a superconductor after high-temperature annealing, and the overall picture indicates that superconductivity in $K_{2-x}Fe_{4+y}Se_5$ originates from the Fe-vacancy order-to-disorder transition. Non-superconducting FeTe with bi-collinear antiferromagnetic order is always considered as the parent compound of FeSe[29,30]. However, a recent TEM study[19] on the non-superconducting $Fe_4Se_5$, which exhibits the $\sqrt{5} \times \sqrt{5}$ Fe-vacancy order and is magnetic, shows that it becomes superconducting after high temperature annealing, meanwhile the Fe-vacancy order disappears.

Our LEED and ARPES results on the MBE grown FeSe thin films show that the $\sqrt{5} \times \sqrt{5}$ Fe-vacancy ordered VO-FeSe is a semiconductor/insulator and can be changed to Fe-vacancy disordered phase(VD-FeSe). One important issue is that whether the VD-FeSe is superconducting or not. The electronic structure of NM-FeSe thin films thicker than 50 UC is almost the same as that measured from FeSe single crystals. It is natural to conclude that the thicker NM-FeSe resembles bulk FeSe with a Tc=8 K. The electronic structure of VD-FeSe phase is found to be electron doped compared with the pristine NM-FeSe. We speculate that the NM-FeSe and the VD-FeSe are both likely superconducting, but their Tc are very low and can't be detected by our ARPES instrument with a lowest detecting temperature of 30 K. Further low temperature STM studies to identify the superconductivity in the NM-FeSe and the VD-FeSe are currently in progress.

In summary, we report the in-situ angle-resolved photoemission spectroscopy study of Fe-vacancy disorder-order transition in FeSe multi-layer thin films grown by molecular beam epitaxy. Low temperature annealed FeSe films are identified to be Fe-vacancy disordered phase and electron doped. Further long-time low temperature anneal can change the Fe-vacancy disordered phase to ordered phase, which is found to be semiconductor/insulator with $\sqrt{5} \times \sqrt{5}$ superstructure and can be reversely changed to disordered phase with high temperature anneal. Our results reveal that the disorder-order transition in FeSe thin films can be simply tuned by vacuum anneal and the $\sqrt{5} \times \sqrt{5}$ Fe-vacancy ordered phase is more likely the parent phase of FeSe.

## ACKNOWLEDGMENTS

We gratefully acknowledge helpful discussions with Dr. X. H. Niu. This work is supported by the Foundation of President of China Academy of Engineering Physics (Grants No. 201501037) and by the National Science Foundation of China (Grants No. 11504341).

———————————————————————————

*sytan4444@163.com

*laixinchun@caep.cn